\documentclass[prb,twocolumn,showpacs,amstext,amsxtra,amssymb,latexsym,showkeys,bbm]{revtex4}
\usepackage{graphicx,epsf}
\usepackage{amsmath,float}
\usepackage{amsfonts}
\usepackage{amsthm}
\usepackage{amssymb}
\usepackage{multirow}
\usepackage{bbm}

\newcommand{\K}{{\boldsymbol K}}
\renewcommand{\k}{{\boldsymbol k}}

\newcommand{\be}{\begin{equation}}
\newcommand{\ee}{\end{equation}}

\newcommand{\D}{{\boldsymbol D}}

\newcommand{\q}{{\boldsymbol q}}

\newcommand{\ep}{\epsilon}

\begin{document}

\title{A semi-Dirac point in the Hofstadter spectrum}

  \author{P. Delplace and G. Montambaux}
\affiliation{Laboratoire de Physique des Solides, CNRS UMR 8502,
Universit\'e Paris-Sud, 91405- Orsay, France}
\date{\today}


\begin{abstract}
The spectrum of tight binding electrons on a square lattice   with half a magnetic flux quantum  per unit cell exhibits two Dirac points at the  band center. We show that, in the presence of an additional uniaxial staggered potential, this pair of Dirac points  may merge into
a single one, with a topological transition towards a gapped phase. At the transition, the spectrum is linear in one direction and quadratic in the other one (a spectrum recently named "hybrid" or "semi-Dirac"). This transition is studied in the framework of a general Hamiltonian describing the merging of Dirac points. The possibility of creating gauge fields for cold atoms in   optical lattices may offer the first opportunity to observe this merging of Dirac points and the hybrid dispersion relation.
\end{abstract}


 \maketitle
 \section{Introduction}
Condensed matter offers the possibility of manipulating the energy spectrum of electrons and modifying their free dispersion relation. Band theory leads to complex dispersion relations, commonly with
{\em quadratic} expansions in the vicinity of peculiar symmetry points. These quadratic expansions are characterized by a tensor of  effective masses,      possibly with positive and negative masses  (near a saddle point). Graphene offers the exciting situation where the dispersion relation at low energy is  {\em linear}, similar to the spectrum of relativistic particles described by the Dirac equation.\cite{Wallace,review} More precisely, the spectrum has the form of two cones (the so-called "Dirac cones" or "Dirac points") in the vicinity of the $\K$ and  $\K'$ points of the reciprocal space.
It has been recently proposed and studied the existence of a {\em hybrid} spectrum, {\em quadratic} in one direction and {\em linear} in the other one, with interesting consequences for the energy spectrum of Landau levels in the presence of a magnetic field.\cite{Dietl} Such a spectrum may appear in a hypothetical graphenelike structure where one ($t'$) of the hopping integrals ($t$) between nearest neighbors is increased.\cite{Hasegawa1,CastroNeto2,Guinea,Montambaux091,Kohmoto09,Montambaux092,Segev} By doing so, the two Dirac points move and, for a critical value of this hopping integral ($t'=2t$), they merge into a single one, with the peculiar hybrid spectrum.

 Other possible systems exhibiting a hybrid spectrum  have been proposed including   the organic conductor $\alpha$-(BEDT-TTF)$_2$I$_3$,
 \cite{Katayama2006,Kobayashi2007,Goerbig2008}
   VO$_2$/TiO$_2$ nanostructures,\cite{Banerjee} or cold atoms trapped in an optical honeycomb lattice.\cite{Zhu,Zhao,Hou2009,Lee09} In the latter example, a recent extensive study has shown the possibility of tuning the position of the Dirac points
 by changing the intensity of the laser fields.\cite{Lee09} In the context of  VO$_2$/TiO$_2$ nanostructures, the hybrid point has been  baptized a "semi-Dirac"
 point,\cite{Banerjee} a name that we will use in this paper.
Quite recently, one of us  has proposed a general framework to study the motion
and the merging of Dirac points in 2D crystals with time-reversal and inversion symmetries,
 within the framework of the following  Hamiltonian \cite{Montambaux091,Montambaux092}

\be {\cal H}_u(\q)= 
\left(
  \begin{array}{cc}
    0 & \Delta+ {q_\parallel^2 \over 2 m^*} -  i c_\perp q_\perp  \\
 \Delta + {q_\parallel^2 \over 2 m^*} +  i c_\perp q_\perp & 0 \\
  \end{array}
\right) \label{newH} \ee
where, by varying the product $m^* \Delta$ from negative to positive values,
a topological transition separates a phase with two Dirac points from a gapped phase,
with a semi-Dirac spectrum at the transition, i.e. when $\Delta=0$.
 This Hamiltonian is universal in the sense that the parameters do not refer to a specific crystal and can be related to
 the microscopic parameters of any 2D system. It has been  shown that in a magnetic field $B$, the Landau
  level spectrum evolves continuously from a $\sqrt{ n B}$ dependence
 to a linear $(n+1/2)B$ behavior, with a  $[(n+1/2)B]^{2/3}$ dependence at the transition.\cite{Dietl,Montambaux091,Montambaux092}

At the moment, there is no straightforward experimental evidence for such merging
of Dirac points in electronic systems. A very interesting alternative
 is  the possibility of fabricating a "crystal" of cold atoms in an artificial optical lattice.
  Unfortunately, the atoms are neutral and are not coupled to a magnetic field.
  A fictitious magnetic field can be induced by a rotation of the sample (then the role of the magnetic field is played by the
Coriolis force \cite{Cooper}), or by effective gauge potentials.\cite{Zoller2003}
 Recently there have  been proposals for the realization of the
 Hofstadter spectrum in a square lattice of cold atoms,\cite{Zoller2003,Gerbier2009}
  using such $U(1)$ gauge potentials, mimicking an external magnetic field. Non-Abelian gauge potentials
  have also been proposed.\cite{Goldman}

The so-called Hofstadter spectrum represents the energy levels of two-dimensional electrons  on a square lattice in a magnetic field, within the nearest-neighbor
  tight binding model.\cite{Hofstadter} The spectrum depends on the value of the magnetic flux $\phi$ through an elementary plaquette of the lattice. For rational values of the reduced magnetic flux, $\varphi=\phi/\phi_0=p/q$, $\phi_0=h/e$ being the flux quantum, the electronic spectrum consists of $q$ sub-bands, leading to the famous fractal structure of the spectrum.

 It is well-known that for a half-flux quantum,
  $\phi=\phi_0/2=h/2e$,  the spectrum exhibits two Dirac cones at  the band center, quite similarly
  to the spectrum of graphene. One may wonder if these two Dirac points can be manipulated
   (moved and merged) by some additional parameter, somehow similarly to the modification
    of the transfer integrals in the honeycomb lattice.
In this paper, we show that an appropriate additional parameter is a staggered potential applied
along one direction. The addition of this uniaxial staggered
potential modifies the position of the Dirac points which may eventually merge
into a "hybrid/semi-Dirac" point for a critical value of the  potential.
The structure of the paper is the following. In section II, we review the basic equations for electrons on a square lattice in a magnetic field and a uniaxial staggered potential.
We discuss the general structure of the resulting Hofstadter spectrum. After a brief  analysis of  the low field spectrum
(sec. III),  we present  our main results in  section IV on  the evolution of the spectrum for a flux in the vicinity
of $\phi_0/2$. After a short analysis of the spectrum near the extrema of the band, we study the spectrum near the center of the
 band ($\ep=0$).
For $\phi_0/2$, it consists of two Dirac cones which merge for a critical value of the staggered potential.
Using the mapping to an effective Hamiltonian near $\ep=0$, we  describe {\em quantitatively} the evolution of the Landau
levels in the vicinity of $\phi_0/2$. We conclude in section V.

\section{Butterfly spectrum with a uniaxial staggered potential}

\begin{figure}[h!]
\includegraphics[width=8cm]{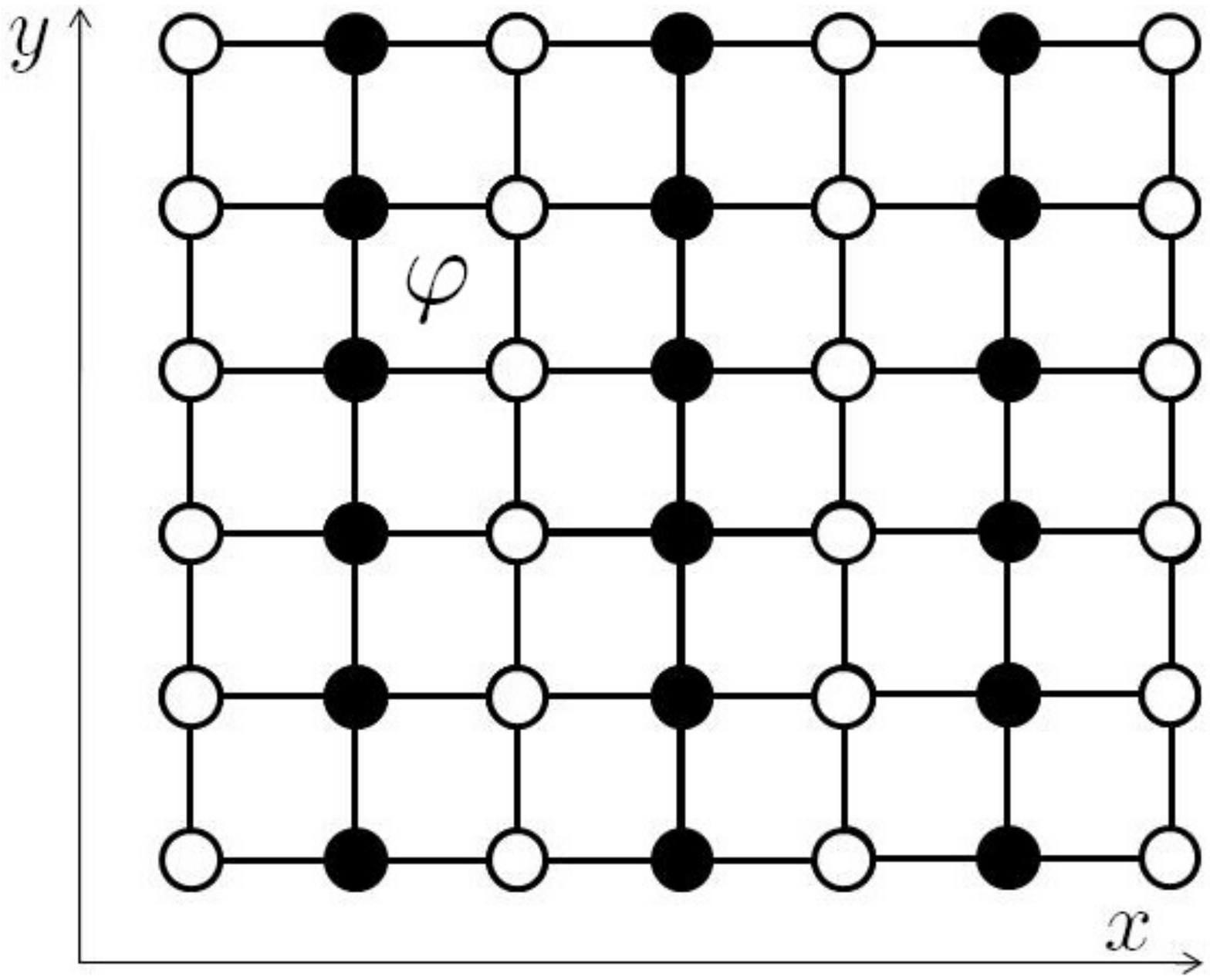}
 \caption{\it Square lattice in a uniform magnetic field and  a uniaxial staggered potential along the $x$ direction. The black and
 white discs represent the on-site potentials $\pm \Delta_s$. $\varphi$ is the dimensionless flux per plaquette.  }
\label{fig.lattice}
\end{figure}

We consider the problem of  tight binding electrons on a square two-dimensional lattice. The  sites are written as $x=m a$ and $y=n a$, $a$ being the lattice spacing. In addition, we apply a uniaxial staggered potential of the form $(-1)^m \Delta_s$ along the $x$ direction, as illustrated in Fig. (\ref{fig.lattice}). The unit cell is made of two atoms having different site energies $\pm \Delta_s$. In the presence of a perpendicular magnetic field $B$, using the Landau gauge where $A_y = B x$, the
Schr\"odinger equation reads

\begin{eqnarray}  E \phi_{m,n} &=& -  t \phi_{m,n+1}e^{-2 i
\pi m \varphi} - t \phi_{m,n-1} e^{2 i \pi m \varphi} \nonumber \\ &&- t
\phi_{m+1,n} - t \phi_{m-1,n}  + (-1)^m \Delta_s \phi_{m,n}  \label{HamE0}\end{eqnarray}
where $\phi_{m,n}$ is the amplitude of the wave function on site $(m,n)$, $t$ is the hopping integral between nearest neighbors and
 $\varphi= e B a^2/h$ is the dimensionless flux through an elementary plaquette of the lattice. Consider the case of a  commensurate flux
 $\varphi=p/q$ where $q$ is even. There are
$q$ inequivalent sites in the unit cell, which is now $q$ times larger along the $x$ direction. We introduce a cell index $l$, so that $m= q l + j$, where $j$ is the position of the site in the unit cell: $j=1,
\cdots, q$.
  Since
there are $q$  sites per unit cell,
 the Brillouin zone in the $x$ direction is $q$ times
smaller and
 Bloch's theorem  implies that

\be \phi_{m,n} = \phi^{(j)}_{l,n}= \psi^{(j)}_{\k} e^{i(k_x l q  + k_y n )a}
\label{Bloch}\ee
with $\k=(k_x,k_y)$ and we have
\be   \psi^{(q+1)}_{\k}= \psi^{(1)}_{\k} \ .  \label{period} \ee
The Schr\"odinger equation now reads
 \begin{eqnarray} E \psi^{(j)}_{\k}&=& -  t
\psi^{(j+1)}_{\k} - t \psi^{(j-1)}_{\k} \\ && - 2 t \psi^{(j)}_{\k} \cos \left(k_y a - 2 \pi j
\varphi\right)+ (-1)^j \Delta_s  \psi^{(j)}_{\k} \nonumber \label{Ham1}\end{eqnarray}
 \begin{eqnarray} E \psi^{(1)}_{\k}&=& -  t
\psi^{(2)}_{\k} - t \psi^{(q)}_{\k}e^{- i q k_x a}  \\ && - 2 t \psi^{(1)}_{\k} \cos \left(k_y a - 2 \pi
\varphi\right)- \Delta_s  \psi^{(1)}_{\k} \nonumber \label{Ham1a}\end{eqnarray}
 \begin{eqnarray} E \psi^{(q)}_{\k}&=& -  t
\psi^{(q-1)}_{\k} - t \psi^{(1)}_{\k}e^{ i q k_x a}  \\ && - 2 t \psi^{(q)}_{\k} \cos \left(k_y a - 2 \pi q
\varphi\right)+ (-1)^q  \Delta_s \psi^{(q)}_{\k}     \nonumber  \label{Ham1b}\end{eqnarray}
This is a $q \times q$ system, with $q$ eigenvalues.
If $q$ is odd,   sites with even and odd $m$ are inequivalent, and the unit cell has a
size $2 q a$ in the $x$ direction. Eqs. (\ref{Ham1},\ref{Ham1a},\ref{Ham1b}) are unchanged (except $q  \rightarrow 2 q$ since  $j= 1, \cdots, 2q$)   and $\psi^{(2 q+1)}_{\k}= \psi^{(1)}_{\k}$.

\begin{figure}[h!]
\includegraphics[width=9cm]{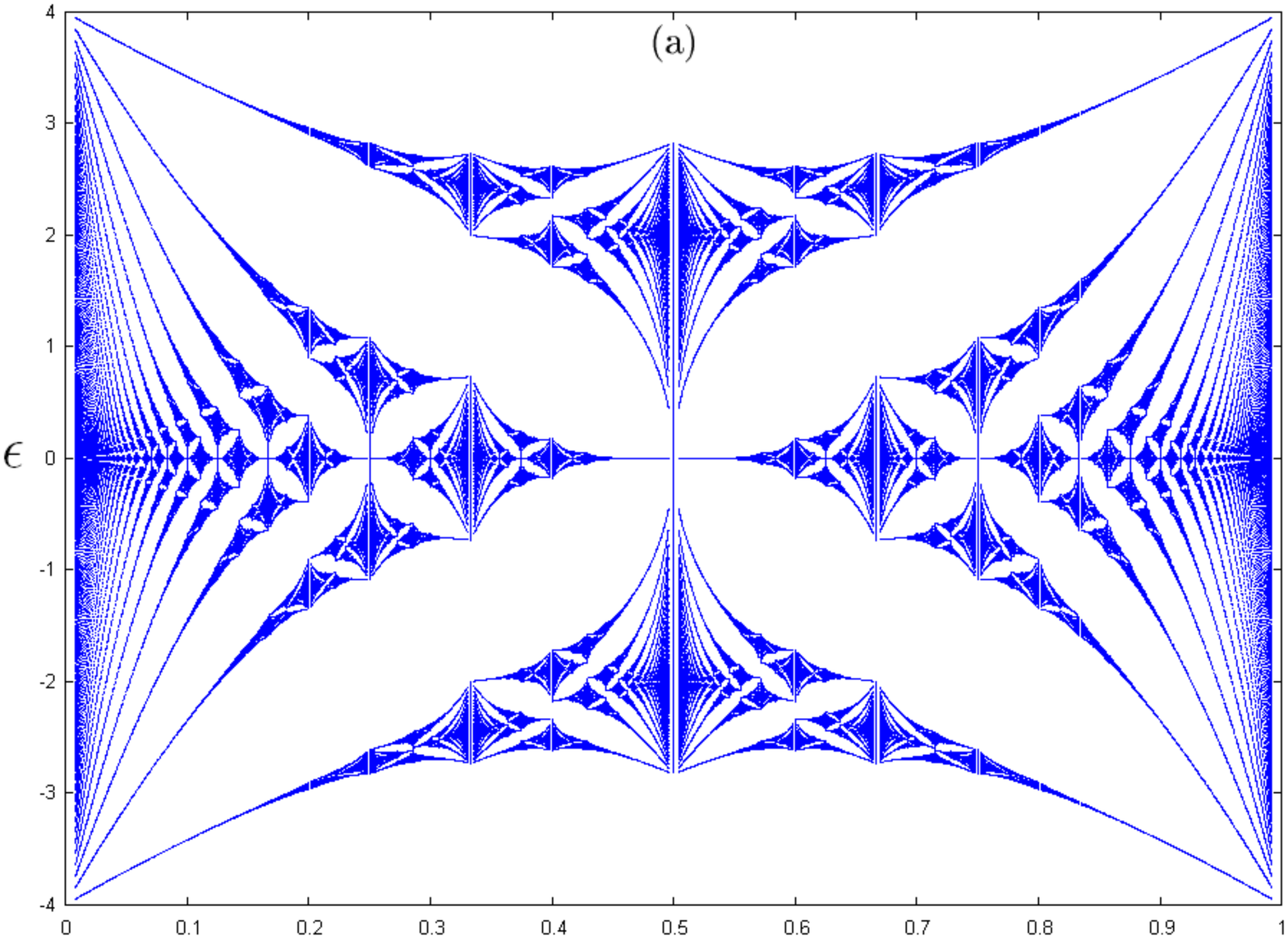}
\includegraphics[width=9cm] {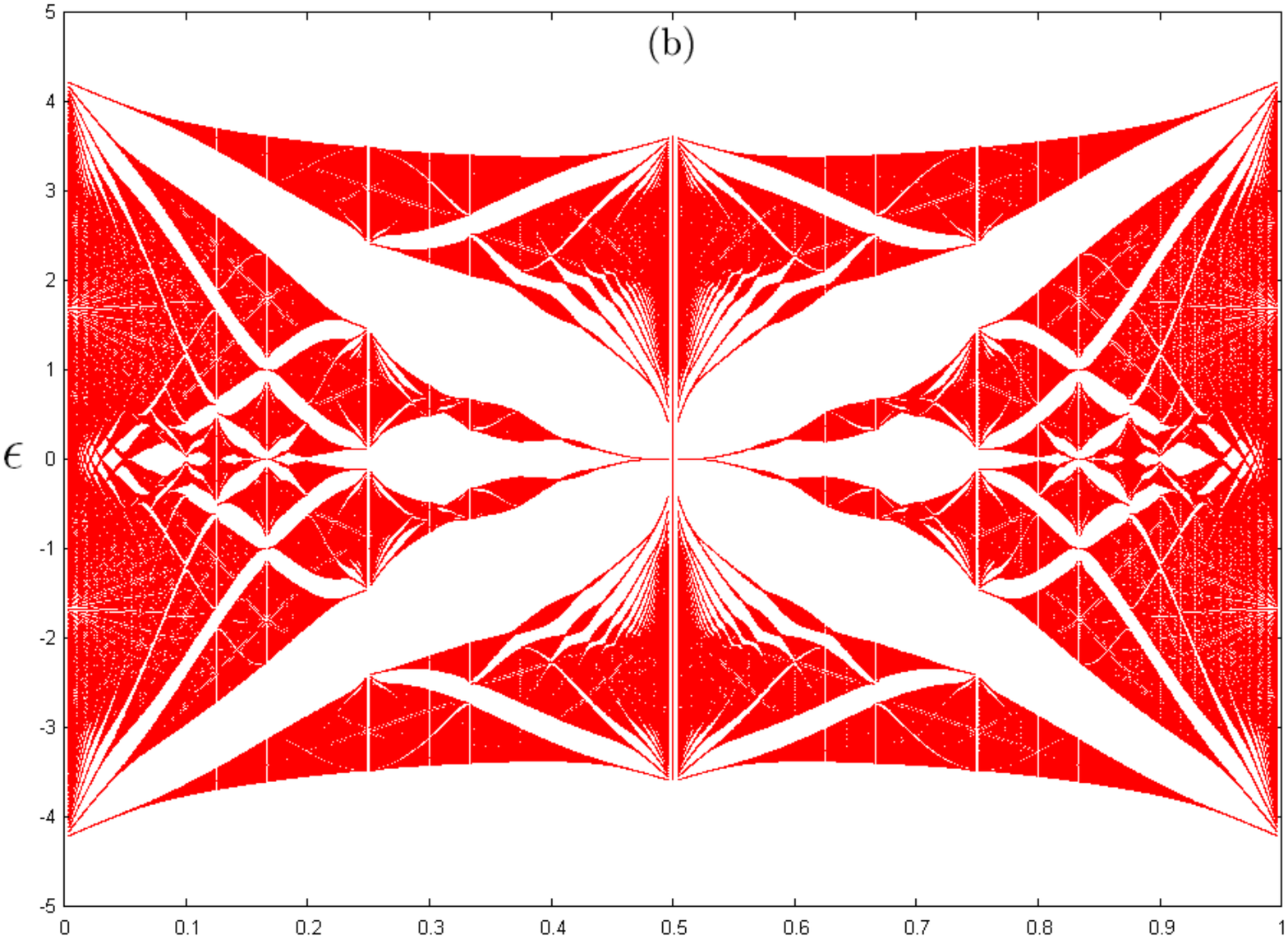}
\includegraphics[width=9cm]{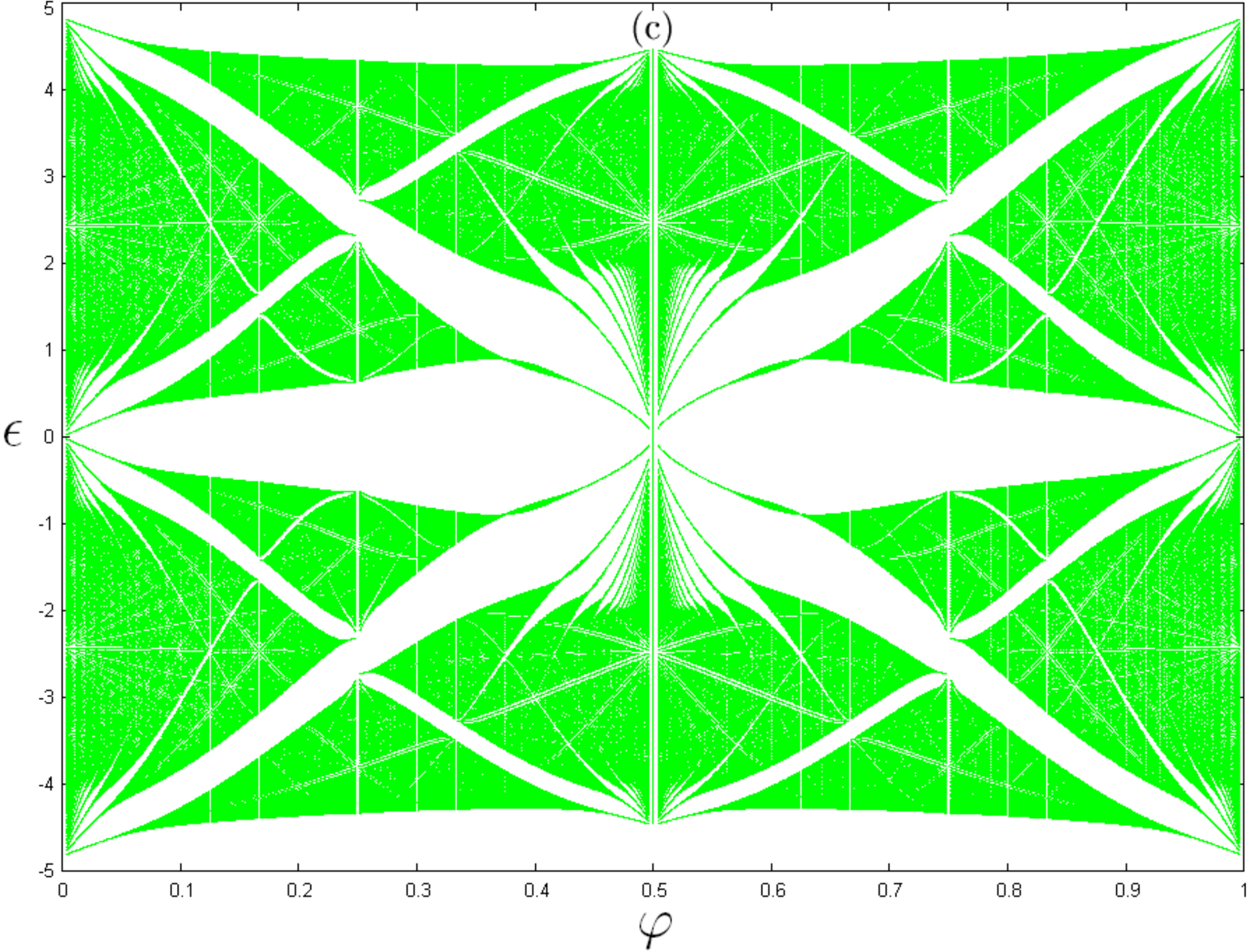}
 \caption{\it Evolution of the Hofstadter spectrum for the  square lattice, in the presence of a uniaxial
  staggered on-site potential characterized by the parameter $r=\Delta_s/(2 t)$.  a) $r=0$,
   this is the usual Hofstadter spectrum.  In the center of the band ($\ep=0$) near $\varphi=1/2$, the energy levels vary as $\sqrt{n f}$, where $f=|\varphi-1/2|$. b) $r=1/2$, the degeneracy of the levels near $\ep=0$, $\varphi=1/2$ has been lifted.  c) Spectrum at the critical point $r=1$. The levels near $\ep=0$, $\varphi=1/2$ vary as $[(n+1/2)f]^{2/3}$. }
\label{fig.hofspectrum.r}
\end{figure}

The figures (\ref{fig.hofspectrum.r}) present the evolution of the spectrum (energy versus magnetic flux) when the staggered potential $\Delta_s$ is increased.
Fig.(\ref{fig.hofspectrum.r}.a) is the familiar Hofstadter spectrum obtained for $\Delta_s=0$.
Fig.(\ref{fig.hofspectrum.r}.b) shows the spectrum for $\Delta_s= t$.
The most striking difference between (\ref{fig.hofspectrum.r}.a)  and (\ref{fig.hofspectrum.r}.b) is that
 many gaps have been filled. This is even more spectacular in Fig.(\ref{fig.hofspectrum.r}.c),
  corresponding to the critical value
$\Delta_s=2 t$ of the staggered potential for which the Dirac points in the  $\varphi=1/2$ dispersion relation merge, as we will see later.\cite{Montambaux90}
This suppression of many gaps is qualitatively understood by the the fact that, when $\Delta_s$ increases,
 there are more and more open classical orbits which are not quantized, as we discuss in   subsection \ref{sect.ZFS}.

The main goal  of our paper is the study of the spectrum near half flux quantum $\varphi=1/2$,
 near the center of the band. For $\Delta_s=0$, this spectrum is quite similar to the low field
 spectrum of graphene : due to the presence of two Dirac points, the spectrum near $\varphi=1/2$
 consists of a series of Landau levels varying as $\sqrt{ n f}$, where $f=|\varphi -1/2|$ is
 the deviation from $\varphi=1/2$ (Fig.\ref{fig.hofspectrum.r}.a).
 These Landau levels are doubly degenerate, due to the two-fold degeneracy of the Dirac spectrum.
  When $\Delta_s$ increases, the degeneracy is progressively lifted (Fig.\ref{fig.hofspectrum.r}.b) and,
   at the critical point $\Delta_s=2$ (Fig.\ref{fig.hofspectrum.r}.c), the levels vary
   as $[(n+1/2) f]^{2/3}$, as we will show in subsection \ref{sect.merging}.
Then, for $\Delta_s >2$, a gap opens and the Landau levels progressively evolve towards a linear variation as $(n+1/2) f$.

We now discuss the different parts of the spectrum, and their evolution when
the staggered potential $\Delta_s$ is increased.

\section{Low field spectrum}

\subsection{Zero field spectrum}
\label{sect.ZFS}

\begin{figure}[h!]
\centerline{ \includegraphics[width=4cm]{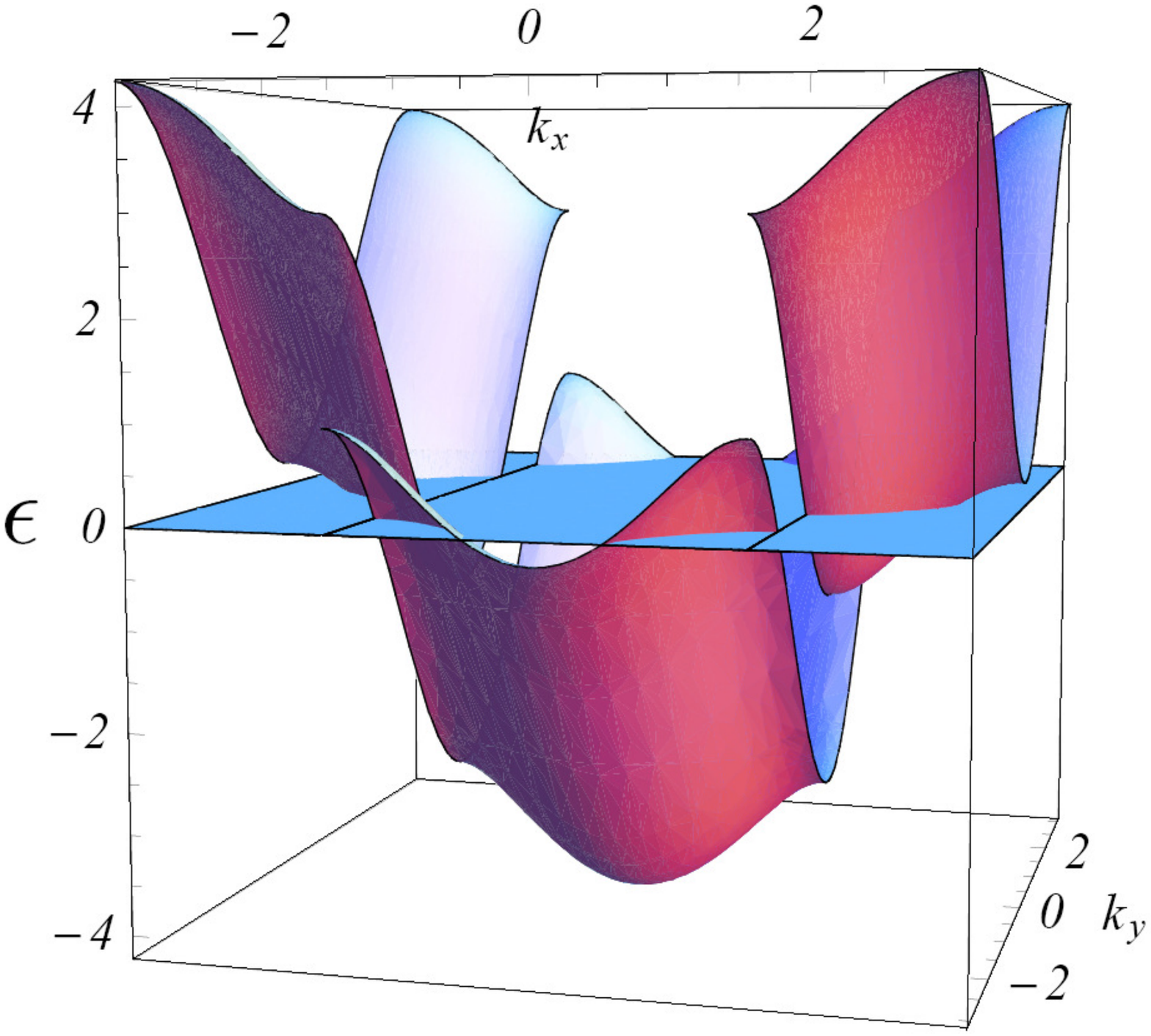}
\includegraphics[width=4cm]{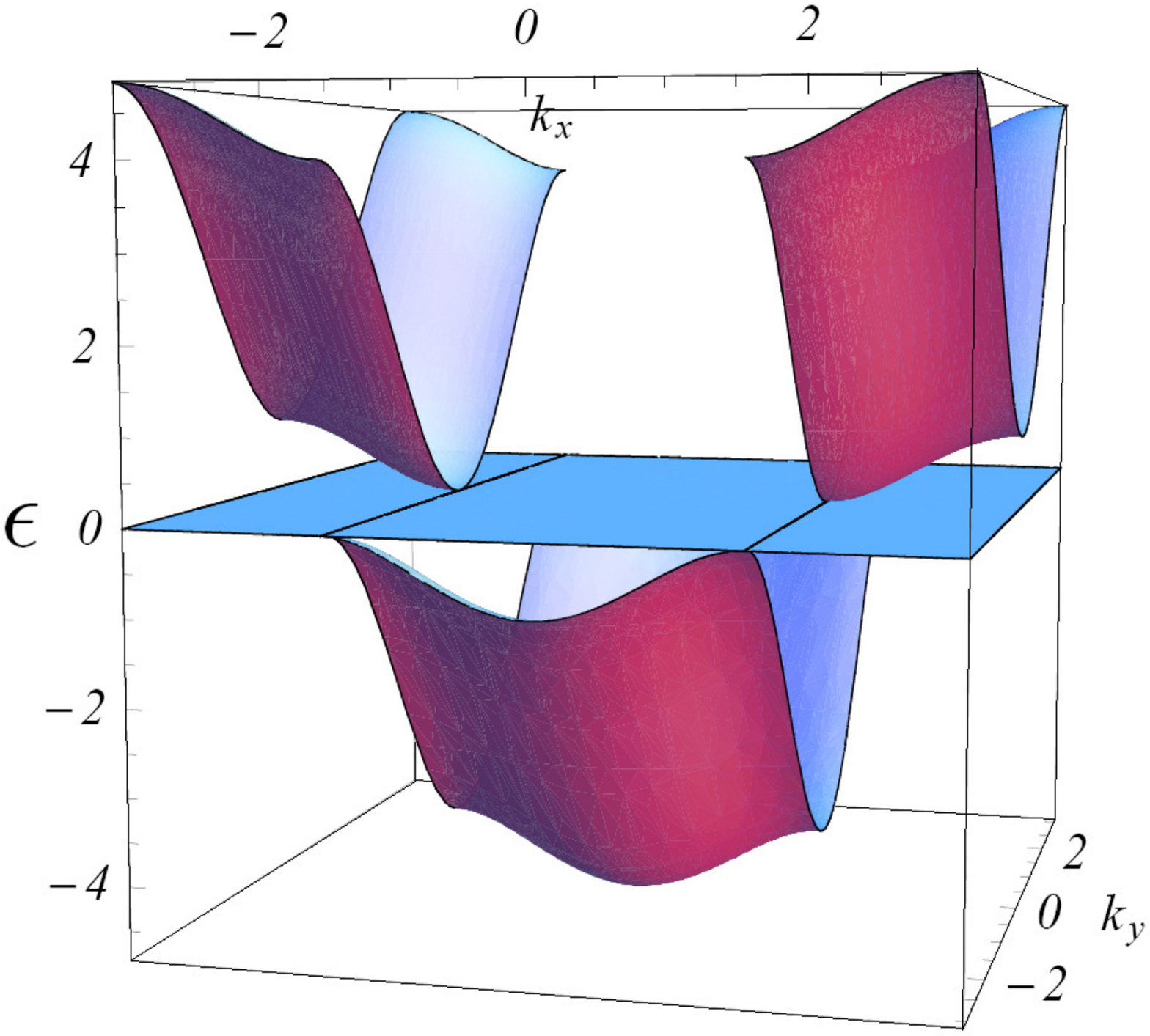}}
\caption{\it  Zero field spectrum $\ep(\k)$ for tight binding electrons on the square lattice with a staggered potential. The two bands overlap when $r <1$.  Left panel: $r=0.5$. Right panel: when $r=1$, the two bands no longer overlap. We have represented the plane $\ep=0$ .}
   \label{fig.zf}
\end{figure}

\begin{figure}[h!]
\centerline{ \includegraphics[width=5cm]{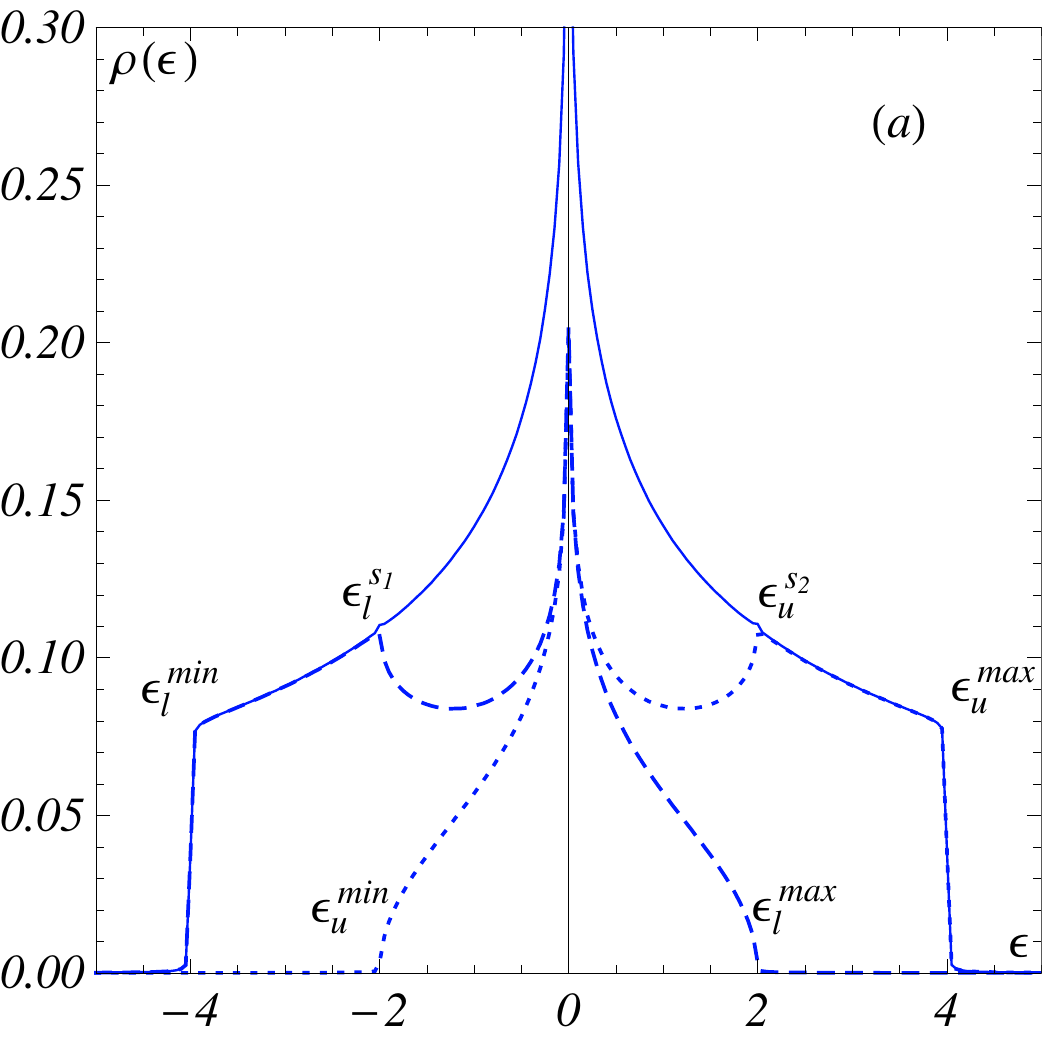}}
\centerline{ \includegraphics[width=5cm]{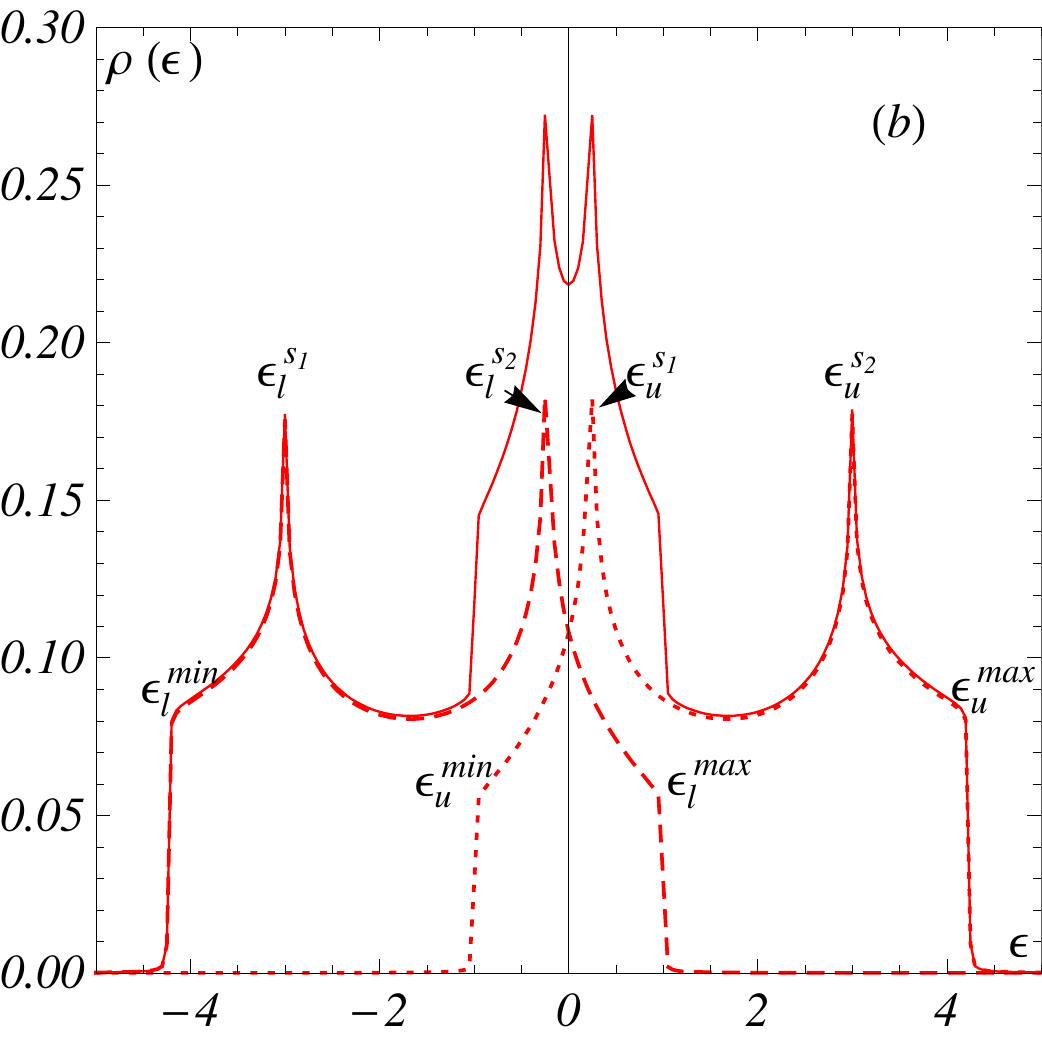}}
\centerline{\includegraphics[width=5cm]{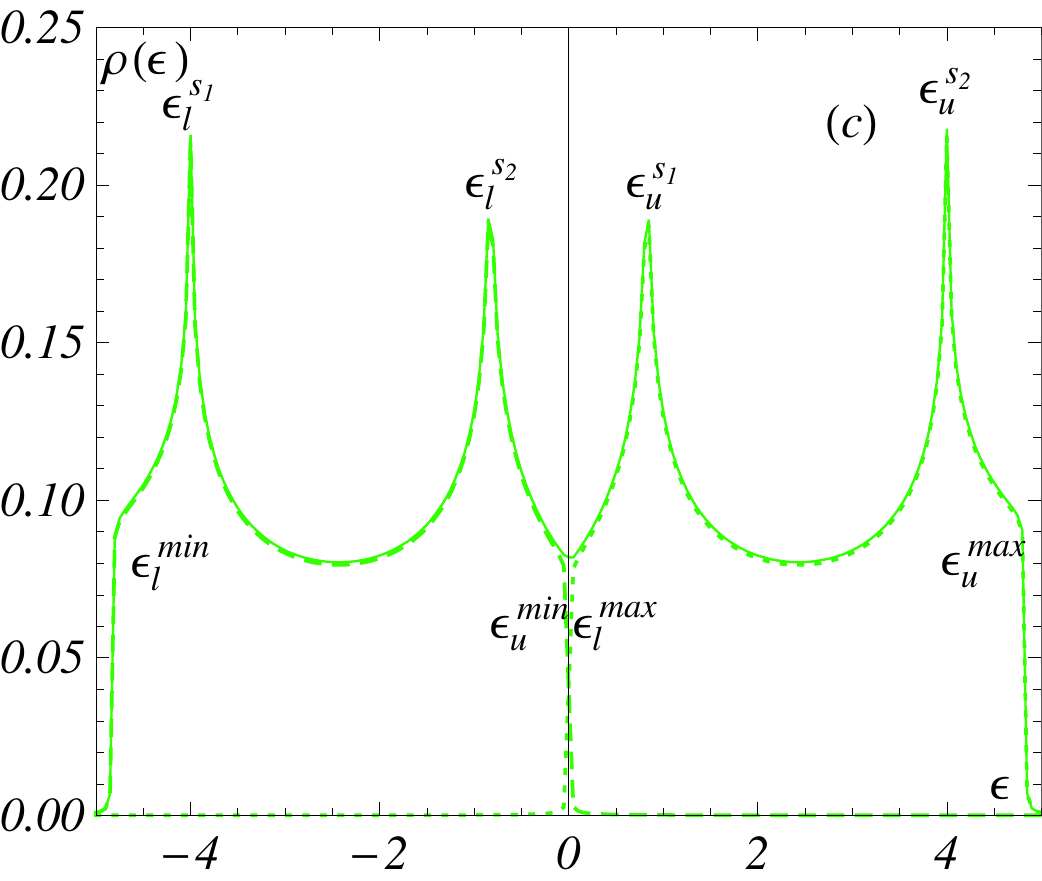}}
\caption{\it{Zero field density of states for tight binding electrons on a square lattice with a
 staggered potential, corresponding to the dispersion relation (\ref{Ezf}), for  $r=0$ (a), $r=0.5$ (b)
  and $r=1$ (c). We have plotted the density of states in the lower band (dashed line), the density of states in the upper band (dotted line) and the total density of states (full line). }}
   \label{fig.dos}
\end{figure}

In zero field, due to the staggered potential, the Hamiltonian has a $2 \times 2$ structure
\be {\cal H}(\k)=-2 \left(
                  \begin{array}{cc}
                    \cos k_y - r & \cos k_x \\
                    \cos k_x & \cos k_y + r \\
                  \end{array}
                \right) \ee
where we introduce the dimensionless parameter $r=\Delta_s/(2t)=\Delta_s/2$, and where we choose from now $t=a=\hbar=e=1$. Therefore the zero field spectrum  is given by
\be \ep(\k)= -2  \cos k_y  \pm 2 \sqrt{\cos^2 k_x + r^2} \label{Ezf}\ee
As shown in Fig.\ref{fig.zf}, the spectrum  consists of two bands which overlap in energy.
The lower band extends from $\ep_l^{min}= -2 -2 \sqrt{1+r^2}$ to $\ep_l^{max}= 2 - 2 r$, with
two saddle points (which yields logarithmic  singularities in the density of states) at energies $\ep_l^{s1}=-2 - 2 r$ and
$\ep_l^{s2}= +2 -2 \sqrt{1+r^2}$. Symmetrically, the upper band extends from $\ep_u^{min}= -2 + 2 r$  to $\ep_u^{max}= 2 +2 \sqrt{1+r^2}$, with
two saddle points (with logarithmic singularities) at energies $\ep_u^{s1}=-2 +2 \sqrt{1+r^2} $ and
$\ep_u^{s2}= 2 + 2 r$. The position of these points is shown  on Fig.(\ref{fig.dos}) on a plot of the density of states.

In each band, the saddle points separate regions with open orbits
(for energies between the two saddle points) and  closed orbits. When $\Delta_s=0$,
all orbits are closed and the total bandwidth for incommensurate fluxes is zero. This can be qualitatively understood on semiclassical grounds: all closed orbits are quantized and magnetic breakdown induces tunneling between orbits which are all quantized.\cite{Thouless} When $\Delta_s$ increases, there are open orbits which are not quantized and the total bandwidth for incommensurate fluxes increases like the number of open orbits.\cite{Thouless} This qualitatively explains why the spectrum is more and more dense when $\Delta_s$ increases.

\subsection{Low field spectrum, at the edges of the band}

 At low magnetic field ($\varphi \ll 1$), the   energy levels near the extrema of the band, are grouped into linear Landau levels (Fig.\ref{fig.hofspectrum.r}). Their spectrum is obtained  from a low energy expansion of the dispersion relation (\ref{Ezf}) near $\k=0$:
\be \ep(\k)=\mp 2(1 + \sqrt{1+r^2}) \pm \left( {k_x^2 \over \sqrt{1+ r^2}}+ k_y^2 \right)+ \cdots \ee
leading to a cyclotron mass $m_0= (1+r^2)^{1/4}/2$ and therefore to a set of Landau levels given by ($B=2 \pi \varphi$ in our notations)

\be \ep_n(\varphi)= \mp (2 + 2 \sqrt{1+ r^2})  \pm {4 \pi \over (1+r)^{1/4}} (n+{1 \over 2}) \varphi + \cdots \ee
which describes properly the evolution of the Landau levels at the extrema of the spectrum (Figs.\ref{fig.hofspectrum.r}).

\subsection{Low field spectrum, center of the  band}

\begin{figure}[h!]
\centerline{\includegraphics[width=10cm]{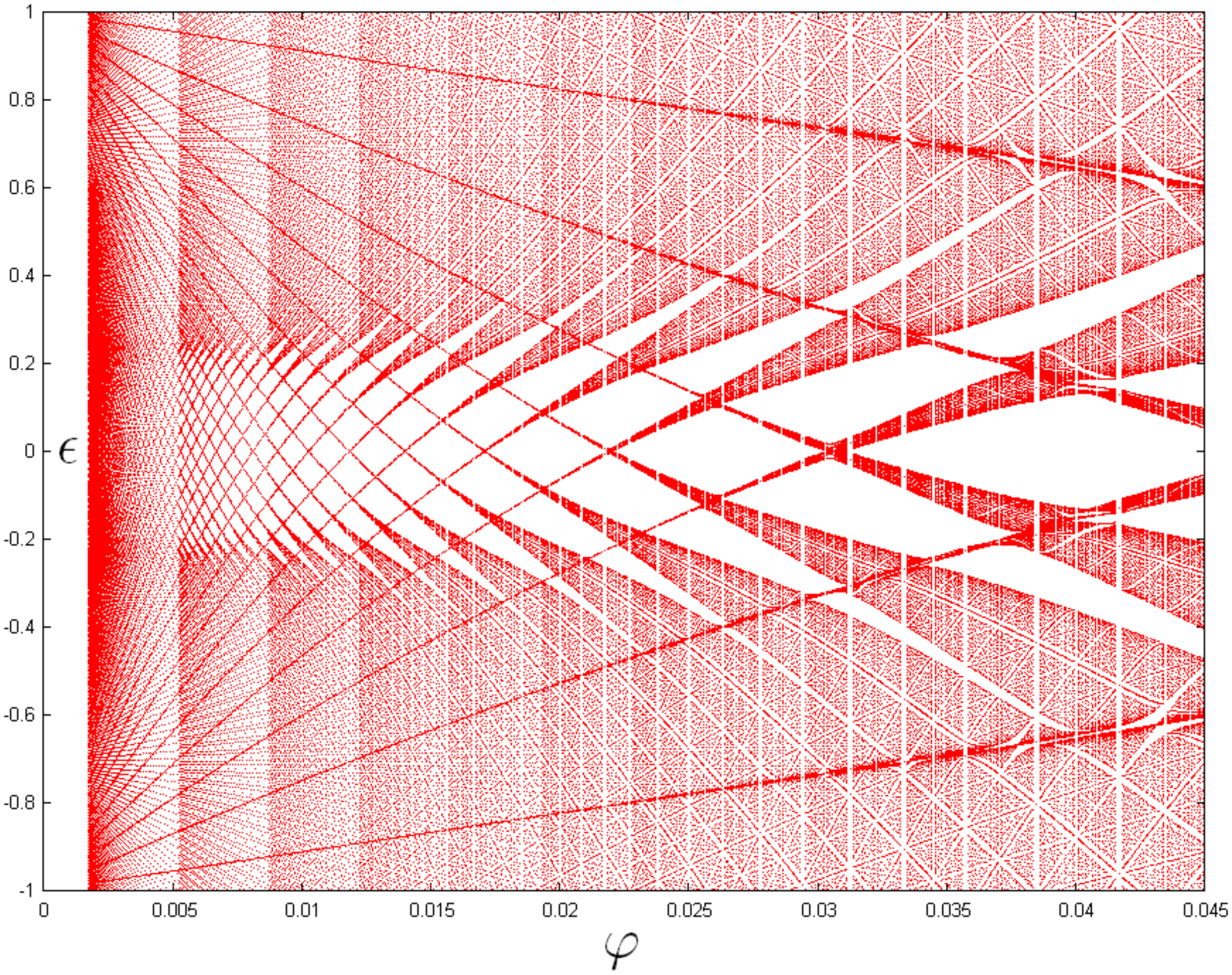}}
\caption{\it  Low field part of the Hofstadter spectrum $\ep(\varphi)$, in the presence of a uniaxial potential, with $r=0.5$, in the energy range $[-1,1]$. The intermixing of the Landau levels is described in the text, and is given analytically by Eq.(\ref{mixedlevels}).}
   \label{fig.cross}
\end{figure}

 Figure(\ref{fig.hofspectrum.r}.b)  shows that, in the middle of the band and for a finite $\Delta_s$,
two set of Landau levels intermix. This is emphasized in the enlargement displayed in Fig.(\ref{fig.cross}). This structure is easily understood by looking at the zero field spectrum, displayed on Fig.(\ref{fig.zf}). When $0 < r < 1$, the spectrum is formed of two overlapping bands. The maximum of the lower band is located at energy $\ep=2 (1 -r)$ and the minimum of the upper band has energy
$\ep=-2 (1 -r)$.  An expansion around these energies gives

\be \ep(\q_\pm)= \mp 2 (1-r) \pm (k_y^2 + {k_x^2 \over r})+ \cdots \ee
where $\q_\pm$ are deviations respectively to the points $(\pi/2,0)$ and $(\pi/2,\pi)$. Therefore the cyclotron mass scales as $m_0= \sqrt{r}/2$, and the two sets of Landau levels are given by

\be \ep_n(\varphi)= \mp 2 (1-r)  \pm  {4 \pi \over \sqrt{r}} (n+{1 \over 2}) \varphi + \cdots  \label{mixedlevels} \ee
When $r=1$, the two sets of Landau levels separate (\ref{fig.hofspectrum.r}.c).

\section{Spectrum near $\varphi=1/2$}

We now concentrate on the region of the butterfly spectrum near $\varphi=1/2$.
Its peculiar structure can be
explained from the spectrum precisely {\em at} $\varphi=1/2$ which plays the role of a zero flux spectrum perturbed by
a small magnetic effective flux $f=|\varphi - 1/2|$, corresponding to  an effective magnetic field $B_f=2 \pi f  $ (the real field being $\pi + B_f$ in our units).
We first describe the low energy part of the spectrum. Then we study the vicinity of the  band center where the spectrum
 consists in two Dirac cones which progressively merge
when the staggered potential increases, until the value $r=1$.

\subsection{Spectrum for $\varphi=1/2$}

\begin{figure}[h!]
\includegraphics[width=4.2cm]{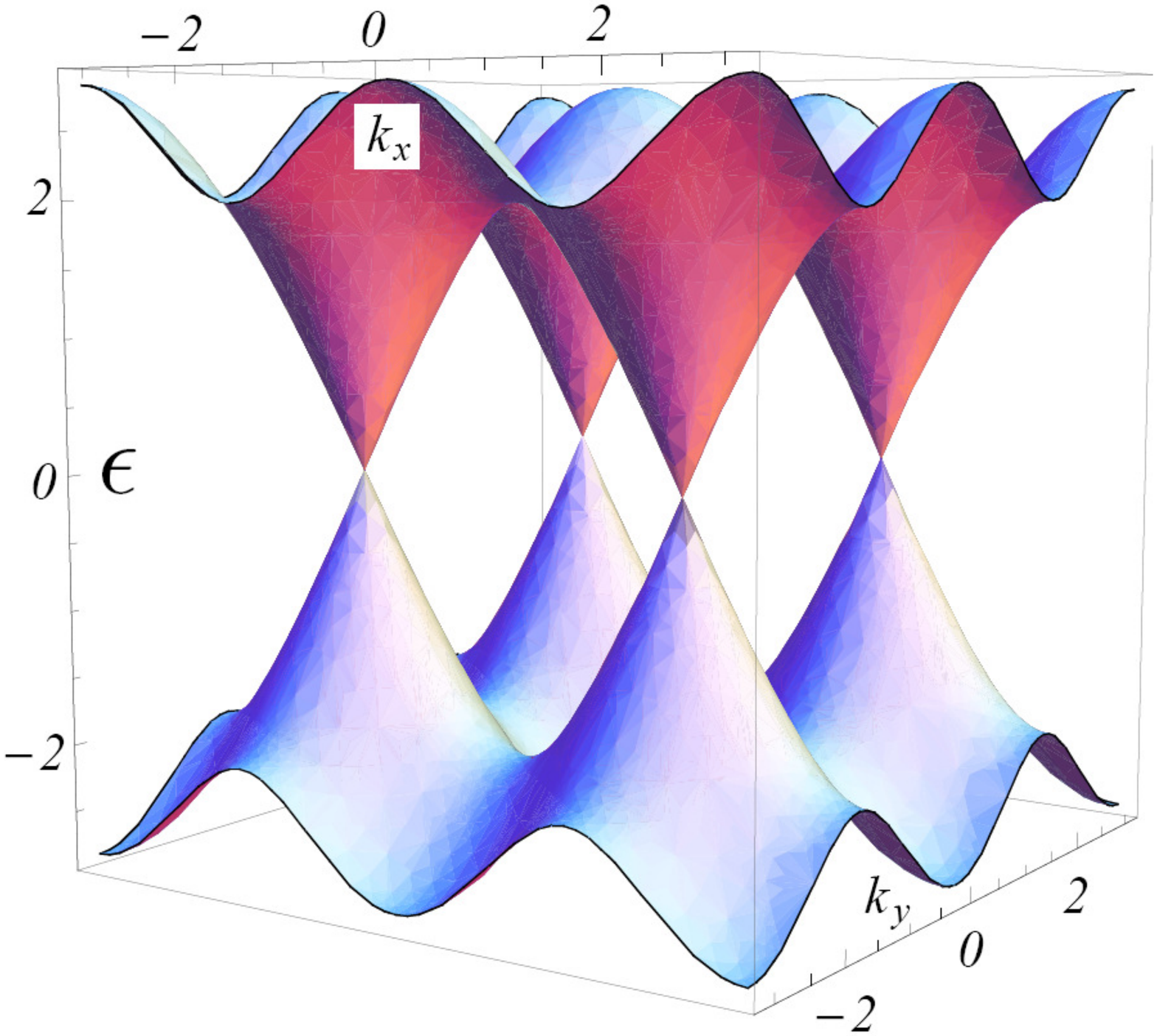}
\includegraphics[width=4.2cm]{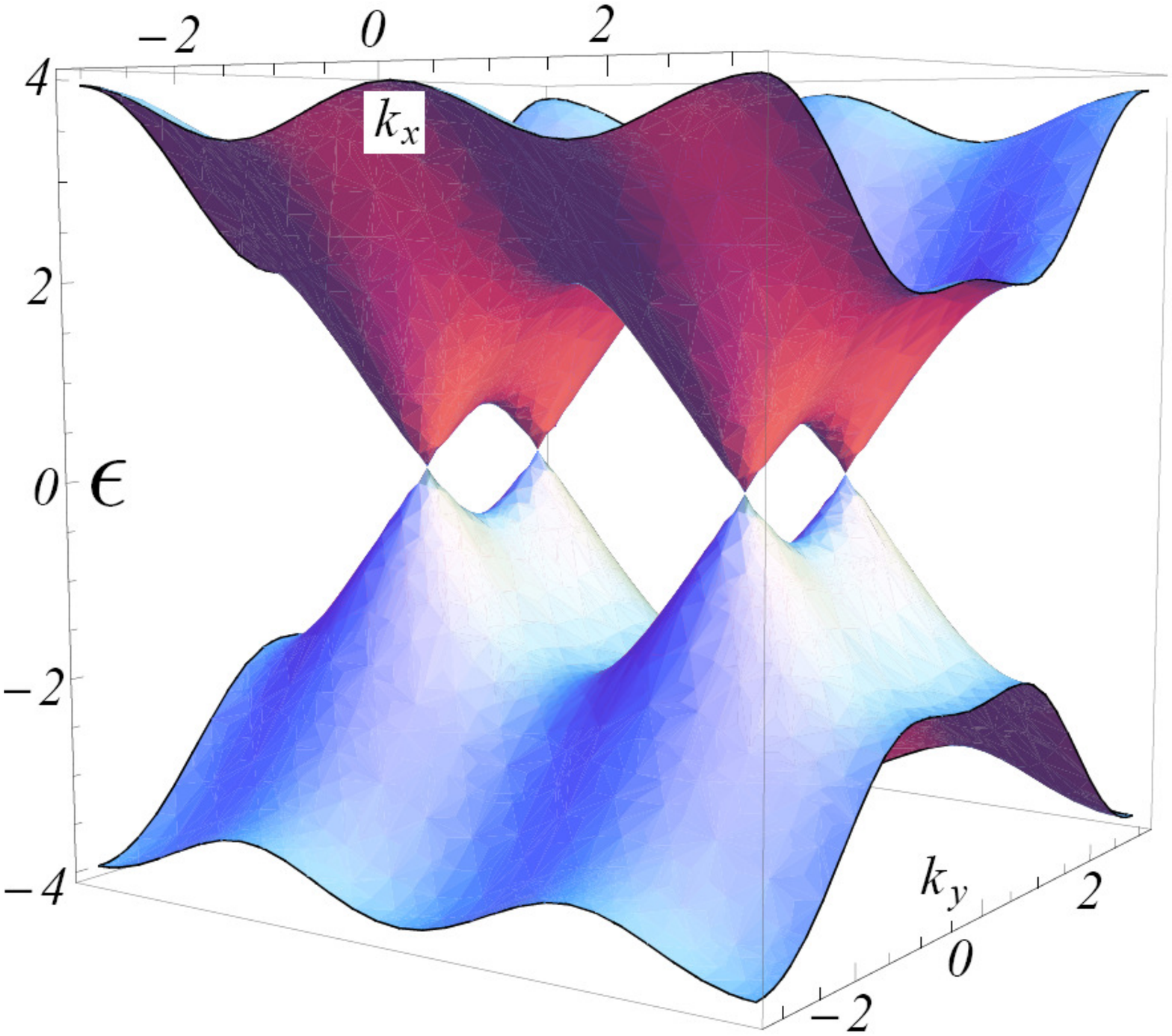}
\includegraphics[width=4.2cm]{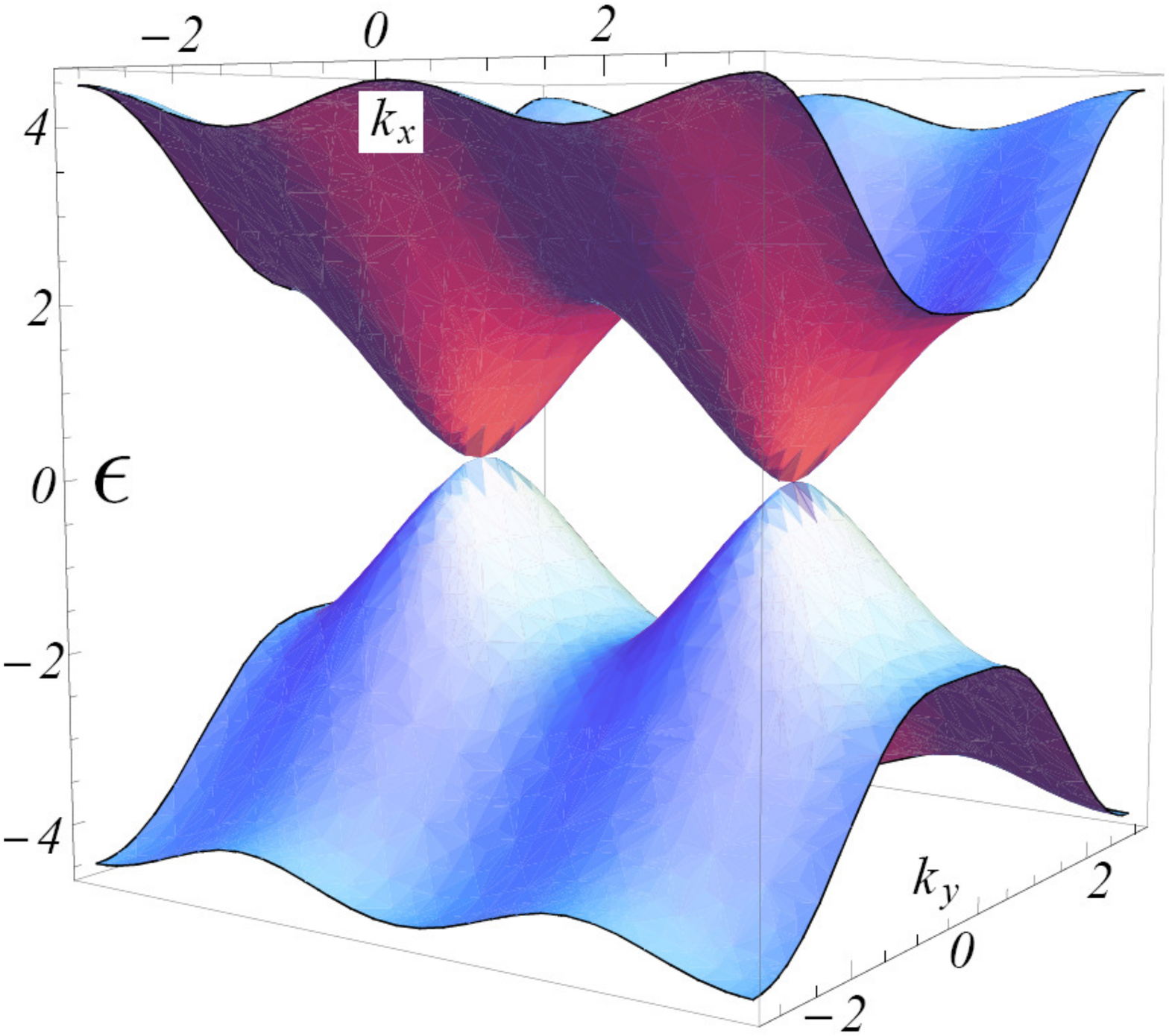}
\includegraphics[width=4.2cm]{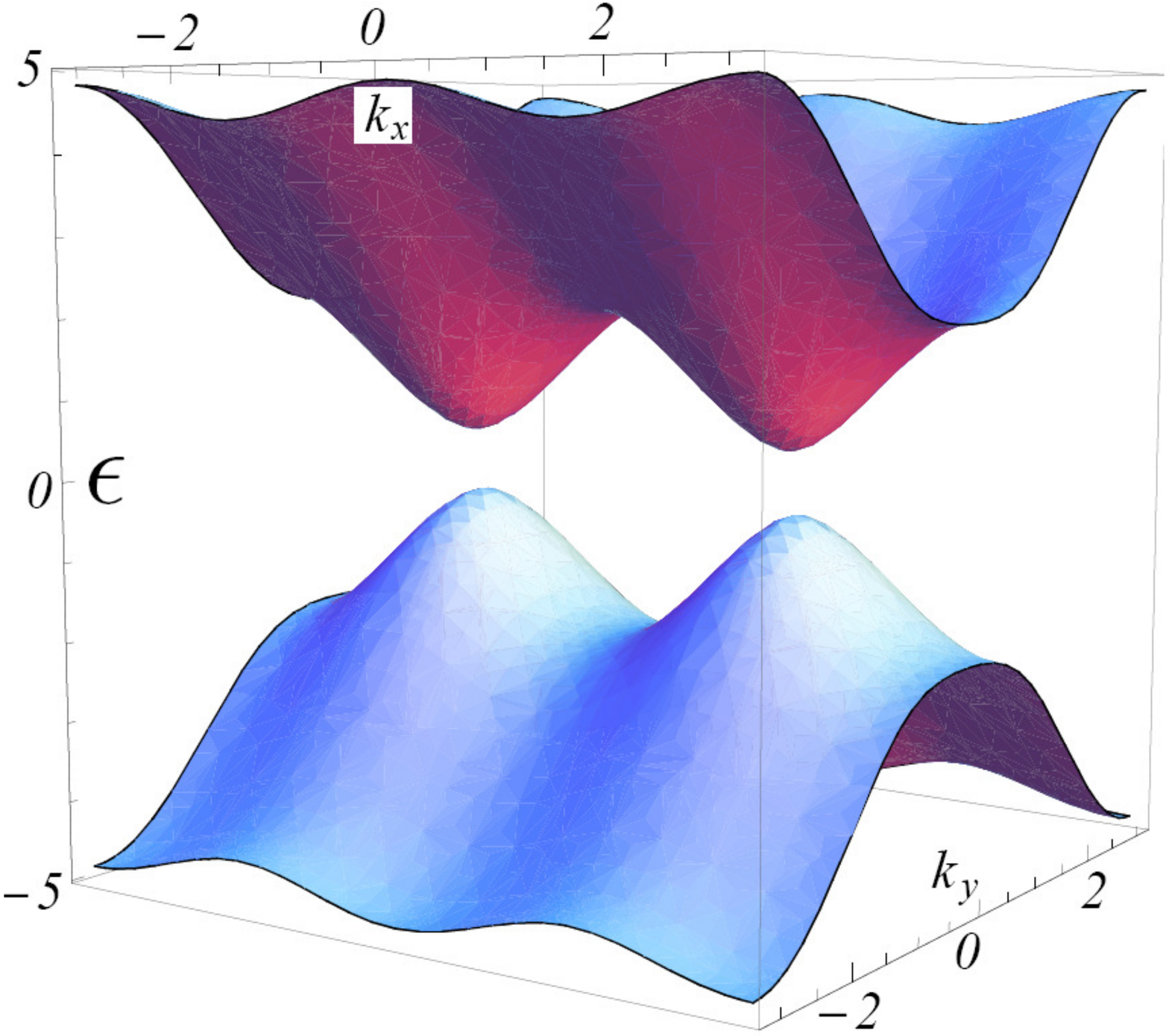}
 \caption{\it Evolution of the electronic spectrum at $\varphi=1/2$, when a uniaxial staggered potential $(-1)^m \Delta_s$ is added. The four plots correspond respectively to $r=\Delta_s/2=0, 0.7, 1$ and $1.2$. Note that that along the $k_x$ direction, we have represented two Brillouin zones, the first zone being $[-\pi/2,\pi/2]$. }
\label{fig.spectrum.r}
\end{figure}

For $\varphi=p/q=1/2$, the Hamiltonian   has the form
\be  {\cal H}(\k) = -   \left(
  \begin{array}{cc}
    2(\cos k_y  - r)  &  1+ e^{-2 i  k_x } \\
     1+ e^{2 i  k_x } & 2(r- \cos k_y)  \\
  \end{array}
\right)\label{Hr1} \ee
with $r= \Delta_s/2$. The energy spectrum   given by
\be \ep(\k)= \pm 2  \sqrt{ \cos^2 k_x  + ( \cos k_y - r)^2} \label{Er1}
\ee
is plotted on Fig.(\ref{fig.spectrum.r}) for different values of $r$. It exhibits a pair of Dirac points and their merging occurs when $r=1$.

\subsection{Spectrum near $\varphi=1/2$, low energy}

An expansion of the dispersion relation
(\ref{Er1}) near its extrema at $\k_m=(0,\pi)$ gives
($\q=\k - \k_m$)

\be \ep(\k_m+\q)=  \mp 2 \sqrt{2+2 r + r^2}  \pm {q_x^2 +(1+r) q_y^2 \over \sqrt{2 +2 r +r^2}} + \cdots \ee
The cyclotron mass $m_0$ is therefore
\be m_0= {\sqrt{2 + 2 r + r^2} \over 2 \sqrt{1+r}} \ee
so that in low "flux" $f$ :

\be \ep_n(f)= \mp 2 \sqrt{2+2 r + r^2}  \pm 4   \pi   \sqrt{{1 +r \over 2 +2 r +r^2}} (n+{1 \over 2}) f + \cdots \ee

\subsection{Spectrum near $\varphi=1/2$, center of the band :   Dirac points and their merging}
\label{sect.merging}

This study is the central point of our paper.
When $r=0$,  the energy vanishes at the two inequivalent points $\D_\xi= (\pi /2, \xi \pi/2)$ where $\xi=\pm 1$ is the valley index.
In the vicinity of these two points, writing $\k=\D_\xi+ \q$, the Hamiltonian can be expanded as
\be  {\cal H}(\D_\xi+ \q) =   2  \left(
  \begin{array}{cc}
    \xi q_y & -i q_x  \\
    i  q_x & -\xi q_y   \\
  \end{array}
\right) \ee
In each valley, the  dispersion relation around $\D_\xi$  is
\be  \ep(\D_\xi+\q) =  2 \xi  q \ ,  \label{Etqa} \ee
as can be seen on Fig.(\ref{fig.spectrum.r}.a).

We now apply a finite uniaxial staggered potential $\Delta_s \neq 0$.  The Dirac points move and are now located at the positions $\D_\xi= (\pi/2, \xi \arccos r) $ (see Fig.\ref{fig.spectrum.r}.b) . A second order expansion around these two points leads to

\be {\cal H}(\D_\xi + \q)=2  \left(
               \begin{array}{cc}
                 \xi \sqrt{1 - r^2}q_y  +  r {q_y^2 \over 2}& -i q_x  \\
                i q_x  & -\xi \sqrt{1 - r^2}q_y  -  r {q_y^2 \over 2} \\
               \end{array}
             \right)
\label{HDexpanded} \ee
with a dispersion relation at small $\q$ of the form
\be
\ep(\D_\xi + \q)= \xi \sqrt{c_x^2 q_x^2 + c_y^2 q_y^2} \label{EDq} \ee
with $c_x=2$ and $c_y= 2 \sqrt{1 - r^2}$.

 When $r$ is increased, the two cones become anisotropic because the velocity along the $y$ direction relating the two cones is reduced (Fig.\ref{fig.spectrum.r}.b).
When $r=1$, the two cones have merged into a single one  at the point $\D_0=(\pi/2,0)$ (Fig.\ref{fig.spectrum.r}.c), and the Hamiltonian becomes

\be {\cal H}_0(\q)=2 \left(
               \begin{array}{cc}
                 {q_y^2 \over 2} & -i q_x     \\
                i q_x   & -{q_y^2 \over 2}  \\
               \end{array}
             \right)
\ee
leading to the "hybrid" dispersion relation

\be \ep(\k)= \pm 2  \sqrt{q_x^2 + q_y^4 / 4} \ .  \ee

Quite remarkably, this dispersion relation is linear in one direction, and quadratic in the other. The study of this semi-Dirac point has been the subject of a series of recent works \cite{Dietl,Hasegawa1,CastroNeto2,Guinea,Montambaux091,Kohmoto09,Montambaux092,Segev,Volovik} and the present problem constitutes a new realization of this merging.

We now wish to describe the physics of the merging, using the effective Hamiltonian

\be {\cal H}=  \left(
               \begin{array}{cc}
               \Delta+  {q_\parallel^2 \over 2 m^*} & -i c_\perp q_\perp     \\
               i  c_\perp q_\perp    & -\Delta -{q_\parallel^2 \over 2 m^*}  \\
               \end{array}
             \right)
\label{Heff}
\ee
with $q_\parallel=q_y$ and $q_\perp= q_x$. Within the rotation ${\cal H}_u = {\cal R} \ {\cal H} \ {{\cal R}^{-1}}$
 where ${\cal R} = 1/\sqrt{2}(\mathbbm{1}-i\sigma_y)$, the Hamiltonian $\cal H$ is identical to the one ${\cal H}_u$ (see Eq.\ref{newH}) that has been recently introduced to describe the merging of Dirac points in a $2D$ crystal with time-reversal and inversion symmetries. When $m^* \Delta >0$, there is a gap in the spectrum. Comparison between (\ref{Heff}) and the expansion of (\ref{Hr1}) near $\D_0$ implies that $\Delta= 2 (r - 1)$. When $m^* \Delta <0$, the spectrum has two Dirac points separated by $2 \sqrt{- 2 m^* \Delta}$ and the  velocity $c_\parallel$ near these Dirac points is  $c_\parallel=\sqrt{- 2 \Delta /m^*}$.

In order to {\em quantitatively} describe the merging of the Dirac points from the Hofstadter spectrum ($r=0$) to the topological transition $(r=1)$, we choose to fix  the velocities $c_\parallel=c_y=2 \sqrt{1 - r^2}$ and $c_\perp=c_x=2$. Then, the mass $m^*$ has to be fixed as $m^*= -2 \Delta/c_\parallel^2=1/(1+r)$. The position of the Dirac points is given by $q_D= \pm 2 \sqrt{{1-r \over 1+ r}}$ slightly different from the real position $q_D=\pm \arccos r$ (see ref. \onlinecite{Montambaux092}, for a discussion on this choice).


Using the universal Hamiltonian (\ref{Heff}) whose properties are known in a magnetic field, we now describe the evolution of the   spectrum when approaching the merging.

\begin{figure}[h!]
\centerline{ \includegraphics[width=6cm]{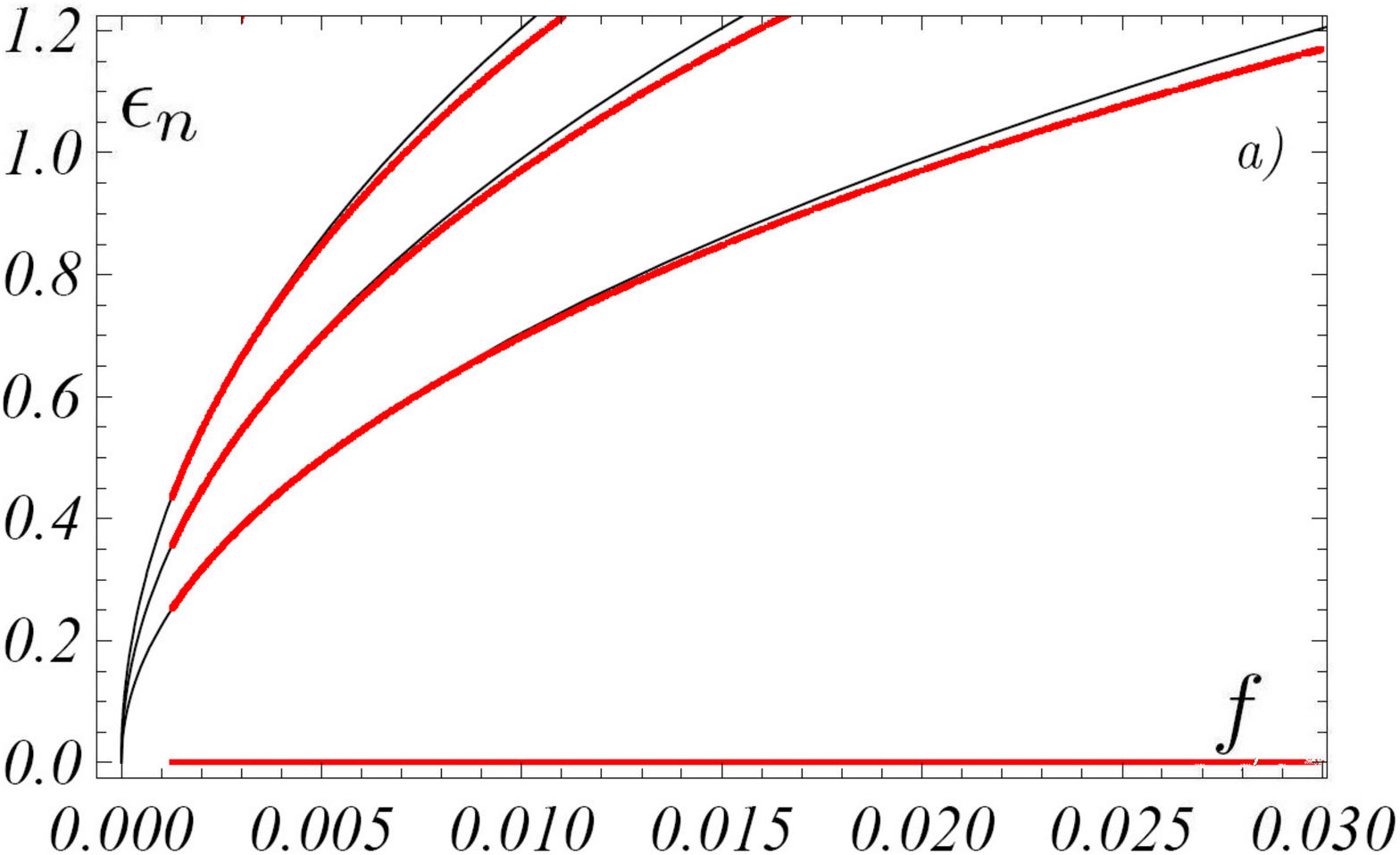}}
\centerline{ \includegraphics[width=6cm]{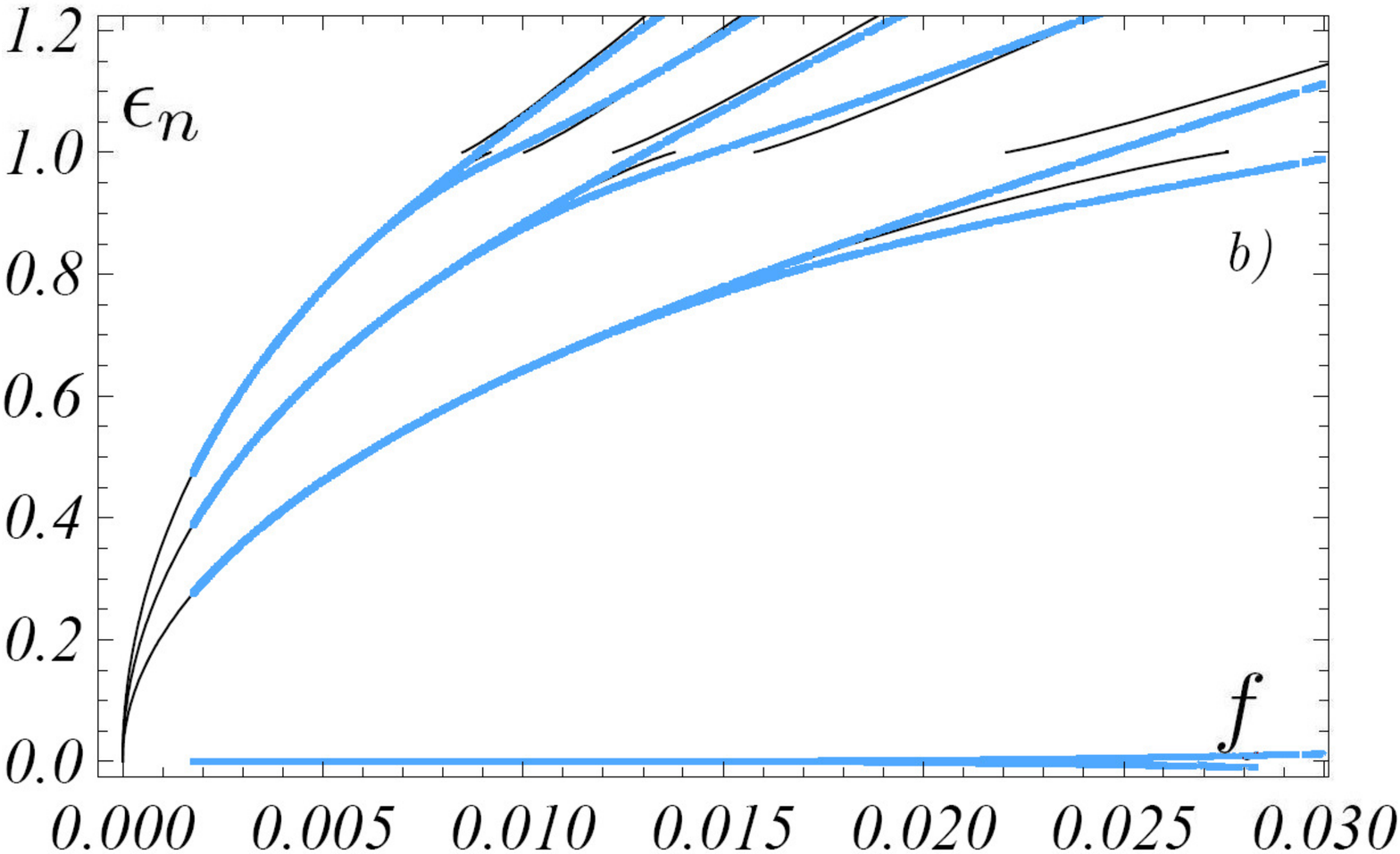}}
\centerline{ \includegraphics[width=6cm]{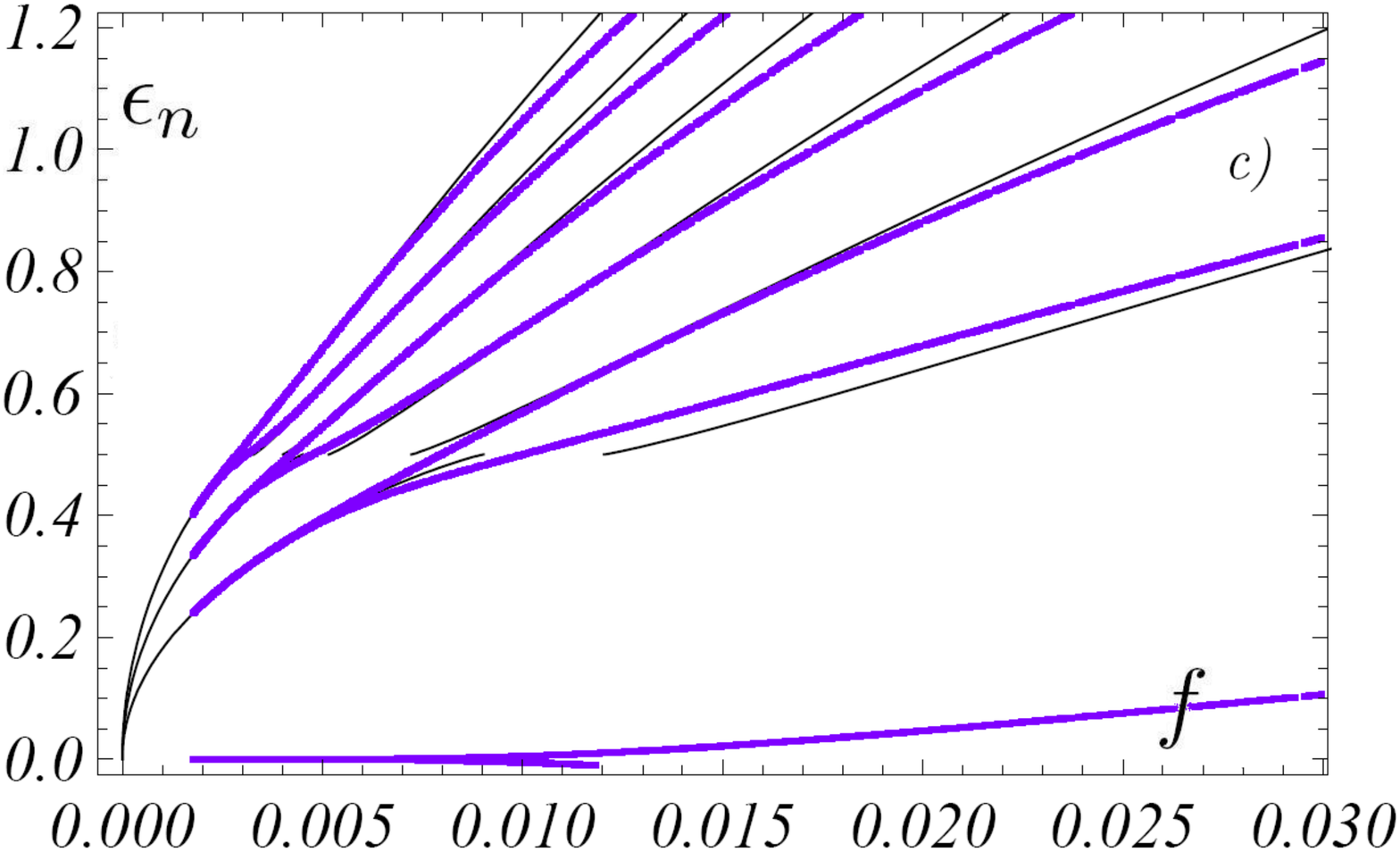}}
\centerline{ \includegraphics[width=6cm]{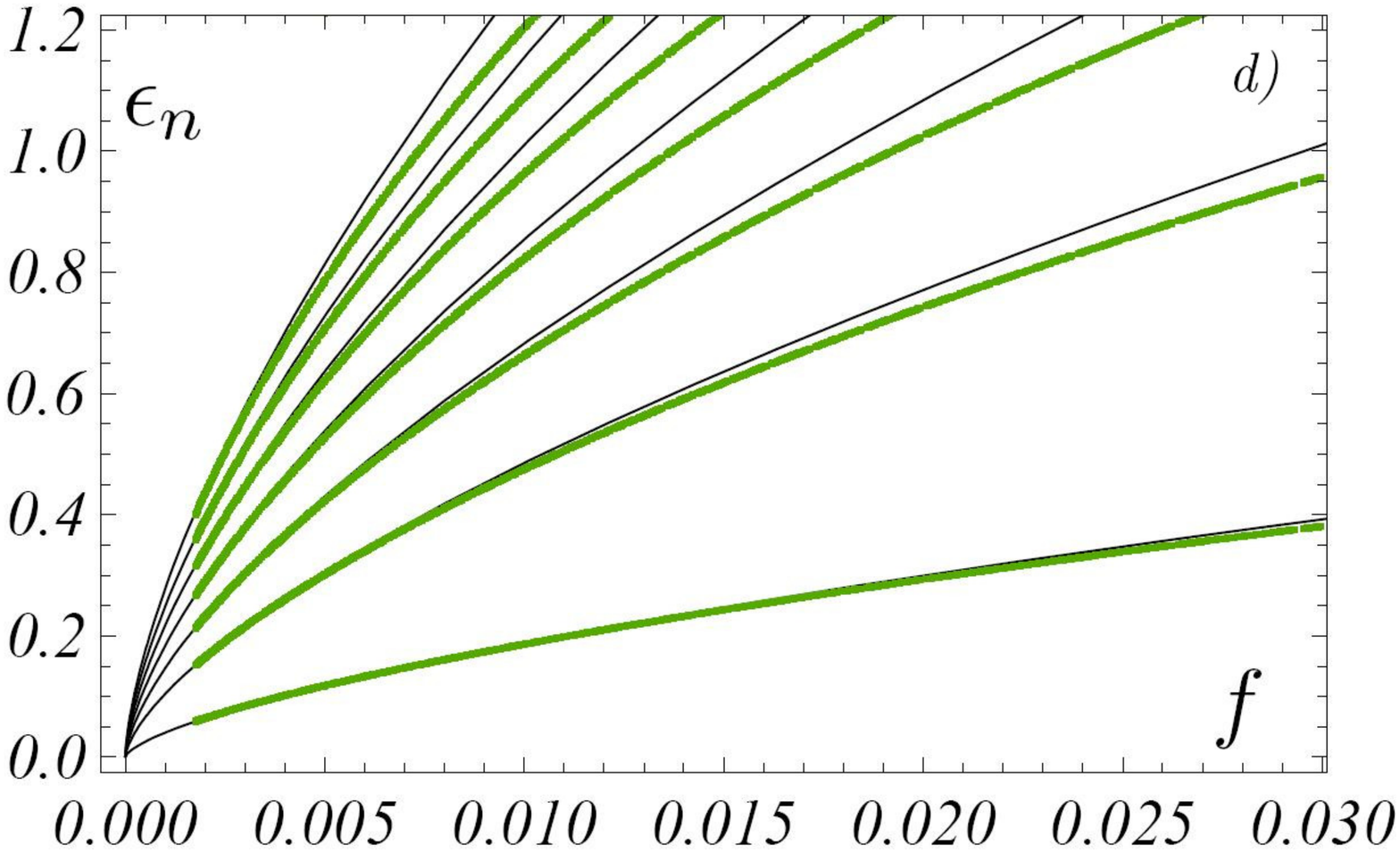}}
 \caption{\it Flux dependence of the energy levels of the tight
binding model on the square lattice, near $\varphi=1/2$, and for an alternate staggered potential characterized by $r=\Delta_s/2$. a) The Hofstadter spectrum ($r=0$). b) $r=.5$, and c) $r=.75$: the two-fold degeneracy
of the levels is progressively removed. d) $r=1$, at the merging of the Dirac cones, the Landau levels vary as $[(n+1/2)f^{2/3}$. The continuous black lines are the results of the semiclassical quantization, which is quite good except, in the vicinity of the energy $\ep= -\Delta= 2 (1 - r)$.  At the merging $r=1$, the energy levels are very well fitted by semiclassical calculation except the ground state for which a numerical factor has been introduced (see text).}
\label{fig.fit1}
\end{figure}

When $r$ is small, the spectrum near $\ep=0$ can still be described as two independent cones with modified velocities $c_\perp=2$ and $c_\parallel=2 \sqrt{1- r^2}$. It is known that in this case,\cite{McClure} the  spectrum is quantized in two-fold degenerate (due to the valley degeneracy) Landau levels with the dispersion relation $\ep_n= \pm \sqrt{2 n c_\parallel c_\perp  e B}$. In our notations, this gives

\be \ep_n(f)= \pm 4 (1-r^2)^{1/4} \sqrt{\pi n f} \label{clurer} \ee
as confirmed on  Figs.(\ref{fig.fit1}.a,b).  When $r$ increases, the domain of validity of this expression is reduced. As seen in Figs.(\ref{fig.hofspectrum.r}.b) and (\ref{fig.fit1}.b), the two-fold   degeneracy of the Landau levels is removed when $f$ is increased (in particular the $n=0$ level), until, for $r=1$, the degeneracy is completely removed, with a new field dependence of the levels.

When $r=0$, the Landau levels are doubly degenerate, due to the valley degeneracy of the two Dirac points. When $r$ increases, the two valleys become coupled and the degeneracy is progressively lifted.
The spectrum in the vicinity of $\varphi=1/2$,
 can be described using semiclassical arguments.
It can be obtained from Bohr-Sommerfeld quantization ${\cal S}= 2 \pi (n+ \gamma) e B$, where ${\cal S}$ is the area of a cyclotron orbit of energy $\ep$ in reciprocal space.

We have to distinguish two regimes (figure \ref{fig.spectrum.r}.b).
In the low energy regime $\ep < -\Delta=2 (1-r)$, this is the area enclosed by {\em one} of the two degenerate iso-energy lines encircling one Dirac point. For the Hamiltonian (\ref{newH}), this area has been calculated in ref. \onlinecite{Montambaux092}. Moreover, it has been argued that due to a Berry phase $\pm \pi$ around each Dirac point, the mismatch factor $\gamma$ is  $0$. Therefore we find

\begin{eqnarray}
&&\sqrt{\ep - \Delta} \left\{(\ep + \Delta) K\left[ R\left({\ep \over \Delta}\right)\right]- \Delta E\left[R\left({\ep \over \Delta}\right)\right]\right\} \nonumber \\
&=& 6 \pi^2\sqrt{{1+r \over 2}} n f \label{SQ1} \end{eqnarray}
where $K(x)$ and $E(x)$ are complete elliptic integrals of the first and second kinds, respectively,\cite{gradstein} and $R(x)=\sqrt{2 x /(x-1)}$.  In the limit $\ep \ll -\Delta$, one recovers (\ref{clurer}).

In the high energy regime $\ep > -\Delta $, that is above the saddle point, the area  enclosed by an iso-energy line encircles {\em the two} Dirac points so that, due to the cancellation of Berry phases, the mismatch factor $\gamma$ is now $1/2$. The calculation of the area ${\cal S(\ep)}$ is this regime gives the semiclassical quantization rule

\begin{eqnarray}
\sqrt{\ep} && \left\{(\ep + \Delta) K\left[ { 1\over R(\ep/\Delta)}\right]- 2 \Delta E\left[ {1 \over R(\ep/\Delta)}\right]\right\}\nonumber \\
&=& 3 \pi^2 \sqrt{1+r}  (n'+{1\over 2}) f \label{SQ2} \end{eqnarray}
In the limit $\Delta \rightarrow 0$, one recovers (\ref{petra}). It is shown on Figs.(\ref{fig.fit1}), that this semiclassical approximation fits very well the numerical calculations, except in the vicinity of the line $\ep=-\Delta$ corresponding to the saddle point separating the two Dirac points (Fig.\ref{fig.spectrum.r}).

To go beyond this semiclassical picture, we may explicitly diagonalize the Hamiltonian (\ref{Heff}) in a magnetic field.  Using the Landau gauge $A_\parallel = B x$, and performing the
Peierls substitution  $q_\parallel - eB x$, one finds that  the eigenvalues $\ep_n$ are solutions of an effective Schr\"odinger equation (see details in refs. \cite{Montambaux091,Montambaux092})
\be \ep_n^2 \psi = \left({m^* \omega_c^2 c_\perp^2 \over 2}\right)^{2/3} (P^2 + (\delta + X^2)^2 -2 X) \psi \label{en2psi} \ee
where $P=\left({2 c_\perp \over m^* \omega_c^2}\right)^{1/3} q_\perp$ and $X= \left({ m^* \omega_c^2 \over 2 c_\perp }\right)^{1/3}   x$ are respectively dimensionless momentum and position ($[X,P]=i$). We have introduced the dimensionless parameter
$ \delta =  \Delta/ \left({m^* \omega_c^2 c_\perp^2 \over 2}\right)^{1/3} $. In our case, this is a unique fonction of the flux $f$ and of the parameter $r$ describing the merging :

\be \delta={(r-1)  \over \pi^{2/3}(1+r)^{1/3}  f^{2/3}} \ee

Eq. (\ref{en2psi}) is an effective  Schr\"odinger equation for  a particle in a double well potential $V(X)=(\delta + X^2)^2 -2 X$.
It has been extensively studied in ref. \onlinecite{Montambaux092}, where its eigenvalues are plotted as a function of the unique parameter $\delta$. Let us recall here its main characteristics. When $\delta <<0$, the two wells of the potential are well separated, one recovers a $\sqrt{n f}$ spectrum of degenerate levels, properly described by eq. (\ref{clurer}), with an excellent fit of the butterfly spectrum near $\varphi=1/2$, $\ep=0$ (Fig.\ref{fig.fit1}.a). When $-\delta$ is decreased, that is when $r$ or $f$ is increasd, the potential barrier between the two wells is reduced, and the tunneling between valleys becomes important. Therefore the two-fold degeneracy of the levels is removed, as seen on Figs.(\ref{fig.fit1}.b,c).

 When $\delta=0$, that is $r=1$, the spectrum
exhibits a single "semi-Dirac" point. The effective potential $V(X)$ is now a quartic potential $V(X)=X^4 - 2 X$. In ref. \onlinecite{Dietl},  a simple WKB quantization argument leads to   $\ep= \pm A (m c_\perp)^{1/3}[(n+1/2) \omega_c]^{2/3}$, where $\omega_c= e B /m^*$ and $A=3^{2/3} \pi / \Gamma(1/4)^{4/3}$.
Using the values of the parameters mentioned above, we expect,
at the critical point

\be \ep= \pm G [(n'+{1 \over 2}) f^{2/3} \label{petra} \ee
with
$G= 2 [36 \pi^5 / \Gamma(1/4)^4]^{1/3}\simeq 7.99 $. This approximation is quite good as soon as $n>1$. However, it has been shown in ref. \onlinecite{Dietl}, that for the ground state, the prefactor has to be multiplied by  a factor $g_0 \simeq 0.808$.\cite{Dietl} Fig.(\ref{fig.fit1}.a) shows a remarkable agreement with the semiclassical calculation.

\section{Conclusion}

It is of great interest to study the physics of Dirac points, their motion and possibly their merging in condensed matter models. In the case of the honeycomb lattice, the motion and merging has been known to be driven by a modification of {\em hopping parameters}, while it is known that a modulated on-site potential opens a gap in the spectrum (like in the case of Boron Nitride), due to parity breaking. Here, we have shown a situation, the Hofstadter problem on a square lattice with half-flux quantum per plaquette $\phi_0/2$,  where the motion and merging of the Dirac points is not due to a change of the hopping integrals, but results from the application of an on-site uniaxial staggered potential. The merging of the Dirac points is discussed within a general hamiltonian, and the structure of the Landau levels near $\phi_0/2$, is explained {\em quantitatively}. There has been a quite recent excitement  about the possible  realization of the Hofstadter spectrum with cold atoms in optical potentials.\cite{Zoller2003,Gerbier2009} The merging of Dirac points described in this paper necessitates the application of a uniaxial staggered potential, which is possible with appropriate laser fields.
\medskip

{\it Acknowledgments - } We acknowledge useful comments from M.-O. Goerbig and F. Pi\'echon.

\end{document}